\documentclass[3p,times,procedia]{elsarticle}
\usepackage{nupha_ecrc}

\volume{00}

\firstpage{1}

\journalname{Nuclear Physics A}

\runauth{Jinjin Pan}

\jid{nupha}

\jnltitlelogo{Nuclear Physics A}

\usepackage{amssymb}

\usepackage[figuresright]{rotating}

\begin{document}

\begin{frontmatter}



\dochead{XXVIIth International Conference on Ultrarelativistic Nucleus-Nucleus Collisions\\ (Quark Matter 2018)}

\title{Balance functions of (un)identified hadrons in Pb--Pb, p--Pb, and pp collisions at the LHC}


\author{Jinjin Pan (on behalf of the ALICE Collaboration)}


\address{Wayne State University, 666 W. Hancock St, Detroit, MI 48201, USA}


\begin{abstract}
In ultrarelativistic heavy-ion collisions, correlations of particles with opposite quantum numbers provide insight into quark production mechanisms and time scales, collective motion, and re-scattering in the hadronic phase. The longitudinal and azimuthal widths of balance functions for charged kaons and charged pions are used to examine the two-wave quark production model proposed to explain quark-antiquark production within the QGP, which predicts a large increase in up and down quark pairs relative to strange quark pairs around the time of hadronization. Balance functions are also analyzed in small collision systems such as p--Pb and pp to study fragmentation effects and possible collective effects in high-multiplicity events. A comprehensive set of balance functions has been measured using the ALICE detector, including results for unidentified hadrons in Pb--Pb and p--Pb collisions at $\sqrt{s_{\rm{NN}}}=5.02$ TeV, for charged pions in Pb--Pb collisions at $\sqrt{s_{\rm{NN}}}=2.76$ TeV, p--Pb collisions at $\sqrt{s_{\rm{NN}}}=5.02$ TeV and pp collisions at $\sqrt{s}=7$ TeV, and for charged kaons in Pb--Pb collisions at $\sqrt{s_{\rm{NN}}}=2.76$ TeV. The first balance function yield results are also presented.
\end{abstract}


\end{frontmatter}


\section{Introduction}

In ultrarelativistic heavy-ion collisions, due to conservation of quantum numbers, a negative balancing charge is produced at approximately the same space--time for each positive general charge. The balance function (BF) is defined as 
\begin{equation}
B(\Delta y,\Delta\varphi) = \frac{1}{2} \{ \frac{\langle N_{+-}\rangle(\Delta y,\Delta\varphi) - \langle N_{++}\rangle(\Delta y,\Delta\varphi)}{\langle N_+\rangle}
 + \frac{\langle N_{-+}\rangle(\Delta y,\Delta\varphi) - \langle N_{--}\rangle(\Delta y,\Delta\varphi)}{\langle N_-\rangle} \},
\label{eq:BFctDeta}
\end{equation}
where $\langle N_{ab}\rangle(\Delta y,\Delta\varphi)$ denotes the average number of pairs per event with charge combination $ab$ ($a,b=+,-$) as a function of two particle rapidity difference $\Delta y$ and azimuthal angle difference $\Delta\varphi$, while $\langle N_a\rangle$ denotes the average number of single particles per event with charge $a$. Thus, the BF
locates general balancing charges in the final state on a statistical basis [1]. The BF is sensitive to multiple phenomena that determine the production of balancing charges and their transport, including two-wave quark production [2], radial flow [3], diffusion [4], quantum statistics [5], and the Coulomb effect.

\section{Analysis details}
\label{}

The results presented in this paper are based on the data sets acquired at the LHC using the ALICE detector [6] during the Pb--Pb runs at $\sqrt{s_{\mathrm{NN}}}=2.76$ TeV in 2010 and at $\sqrt{s_{\mathrm{NN}}}=5.02$ TeV in 2015, the p--Pb runs at $\sqrt{s_{\mathrm{NN}}}=5.02$ TeV in 2013 and 2016, and the pp runs at $\sqrt{s}=7$ TeV in 2010. The acquisition of events is triggered with a minimum bias trigger using the V0 detector, which requires a coincidence of hits in both V$0$A and V$0$C detectors in Pb--Pb and p--Pb collisions, while in pp collisions it requires at least one hit in the Silicon Pixel Detector (SPD) or in one of V$0$A and V$0$C. Background events, such as generated by beam--gas interactions, are eliminated to a negligible level ($<0.1 \%$) by comparing the signal arrival time between V$0$A  and V$0$C with 1 ns resolution. In total, with all the selection criteria applied, about $14 \times 10^{6}$, $82 \times 10^{6}$, $100 \times 10^{6}$, $620 \times 10^{6}$ and $240 \times 10^{6}$ events are analyzed for Pb--Pb at 2.76 TeV and 5.02 TeV, p--Pb in 2013 and 2016, and pp collisions, respectively. The particle identification is achieved by using combined information from the Time Projection Chamber (TPC) and the Time-Of-Flight (TOF) detector, with purity of both charged kaons (${\rm K}^{\pm}$) and charged pions ($\pi^{\pm}$) better than 96\%. For unidentified hadrons (${\rm h}^{\pm}$), the pseudorapidity selection is $|\eta| \le 0.8$, while for ${\rm K}^{\pm}$ and $\pi^{\pm}$, the rapidity selection is $|y|\le 0.8$. Detection efficiency and acceptance are fully corrected using a weight technique [7].

\section{Results}
\label{}

\begin{figure}[h!]
 \includegraphics[width=0.33\linewidth]{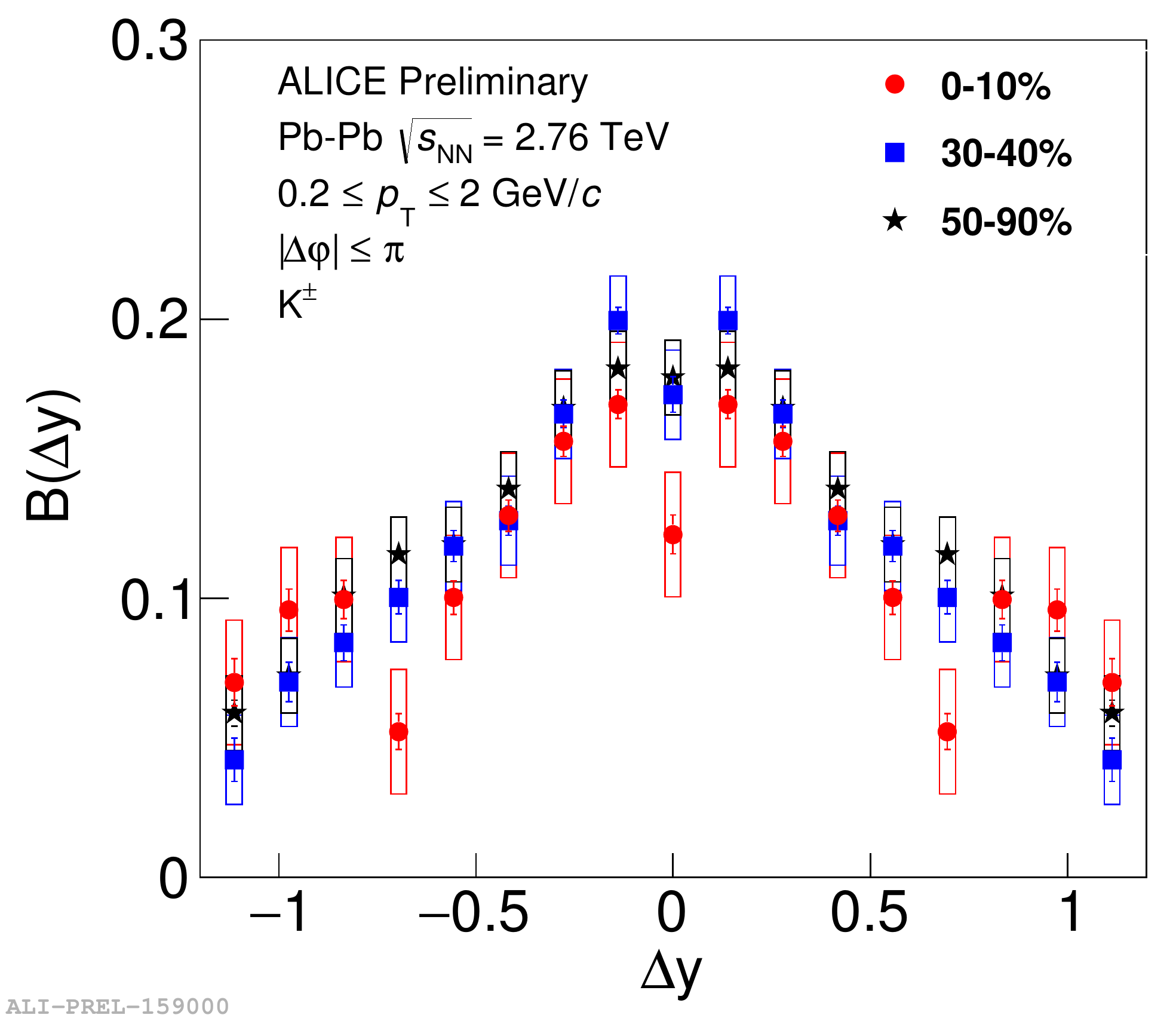}
 \includegraphics[width=0.33\linewidth]{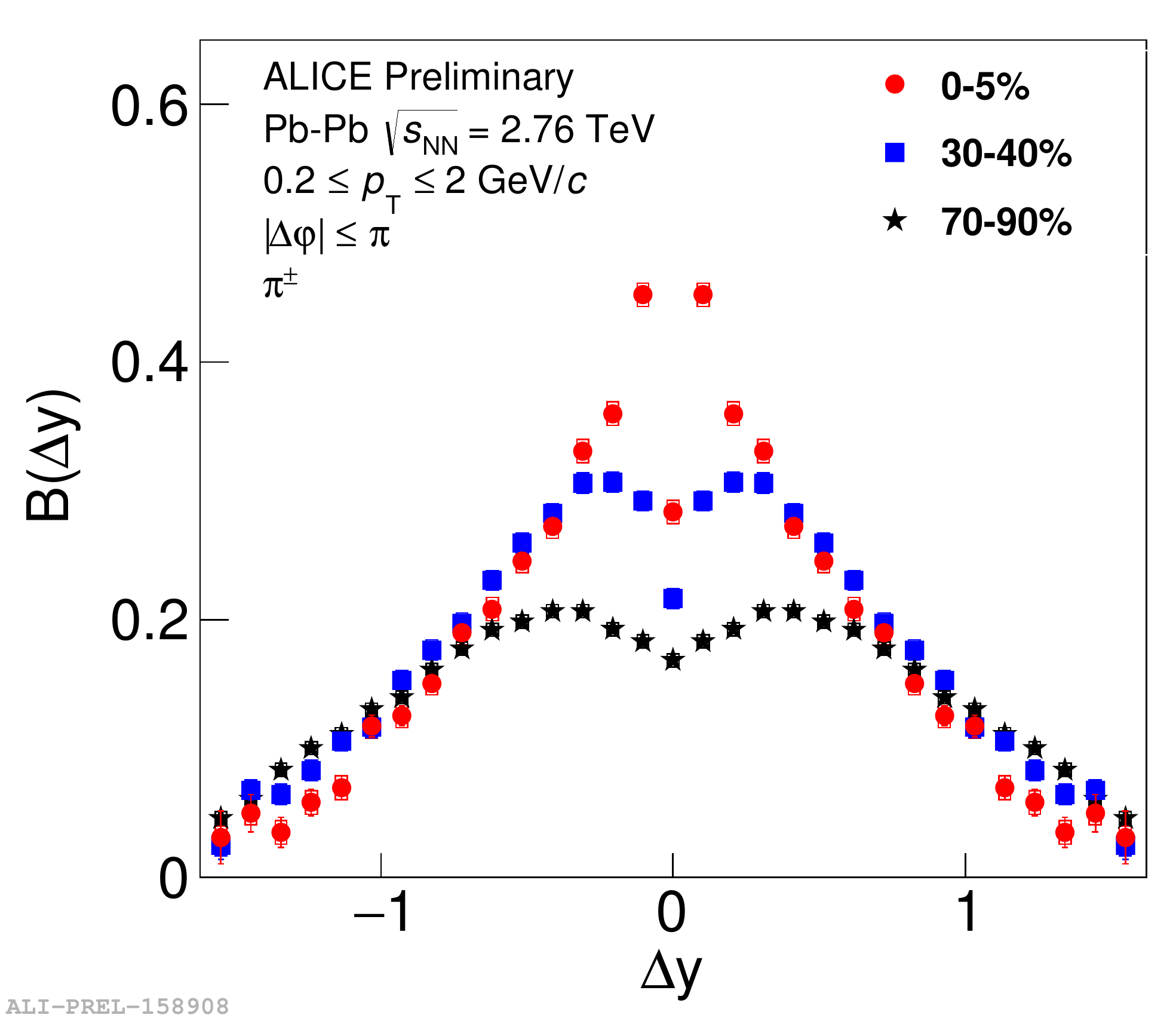}
 \includegraphics[width=0.33\linewidth]{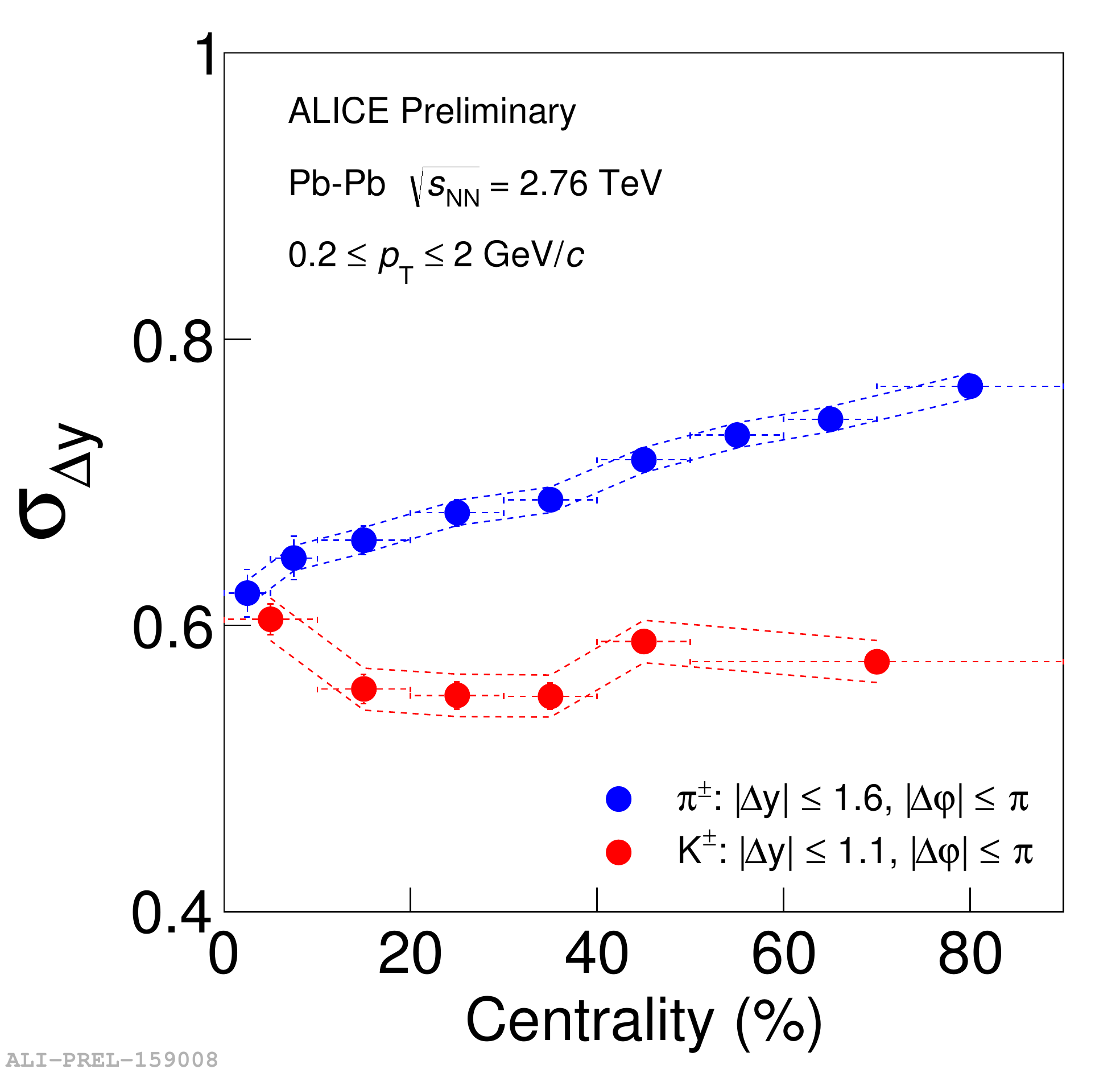}
 \caption{The BF as a function of $\Delta y$ for ${\rm K}^{\pm}$ (left) and $\pi^{\pm}$ (middle) in selected centralities, and their RMS widths as a function of centrality (right) in Pb--Pb collisions at $\sqrt{s_{\mathrm{NN}}}=2.76$ TeV.}
  \label{fig:1d_PID_BF_dy_PbPb}
\end{figure}
\begin{figure}[h!]
  \includegraphics[width=0.33\linewidth]{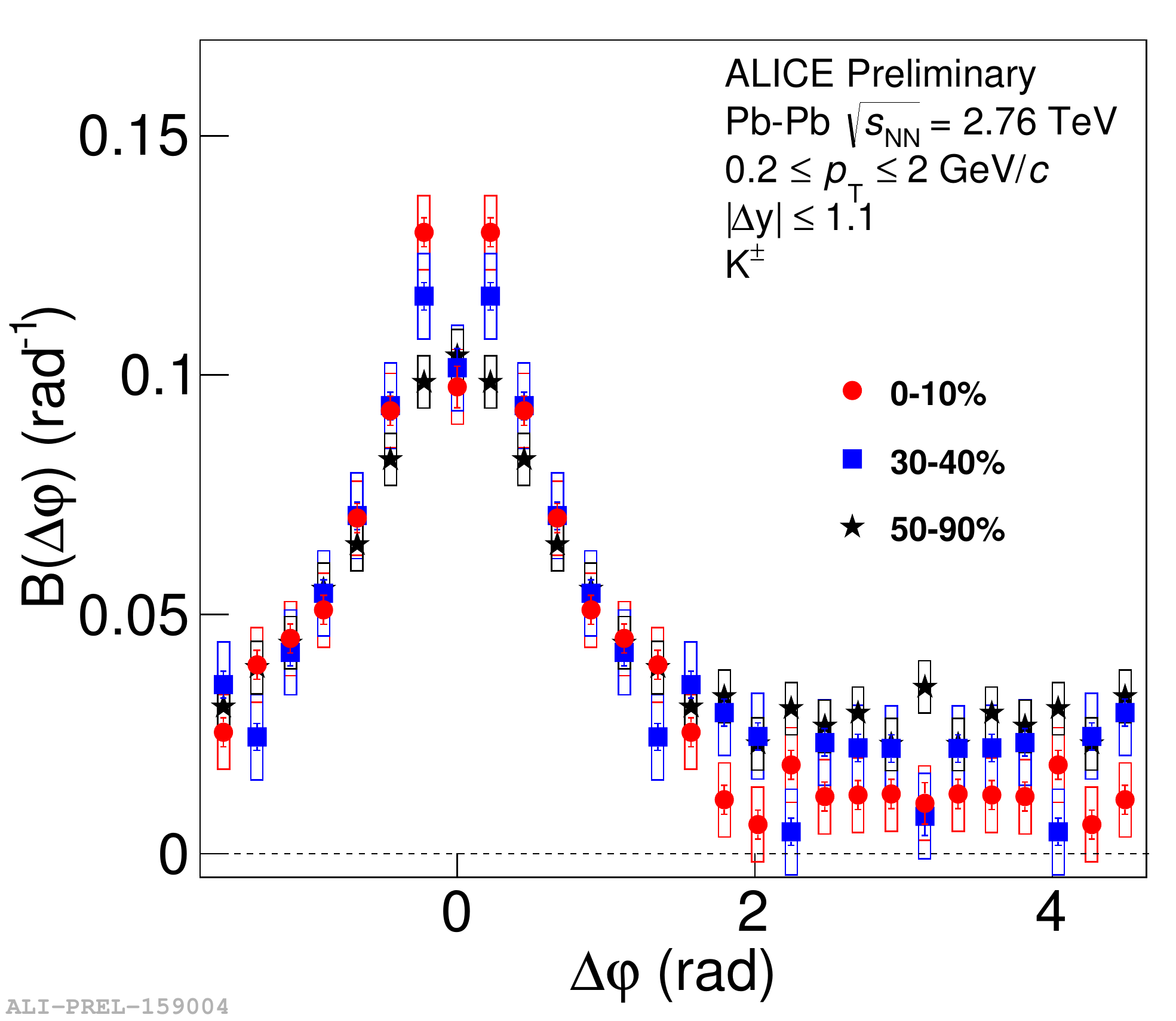}
 \includegraphics[width=0.33\linewidth]{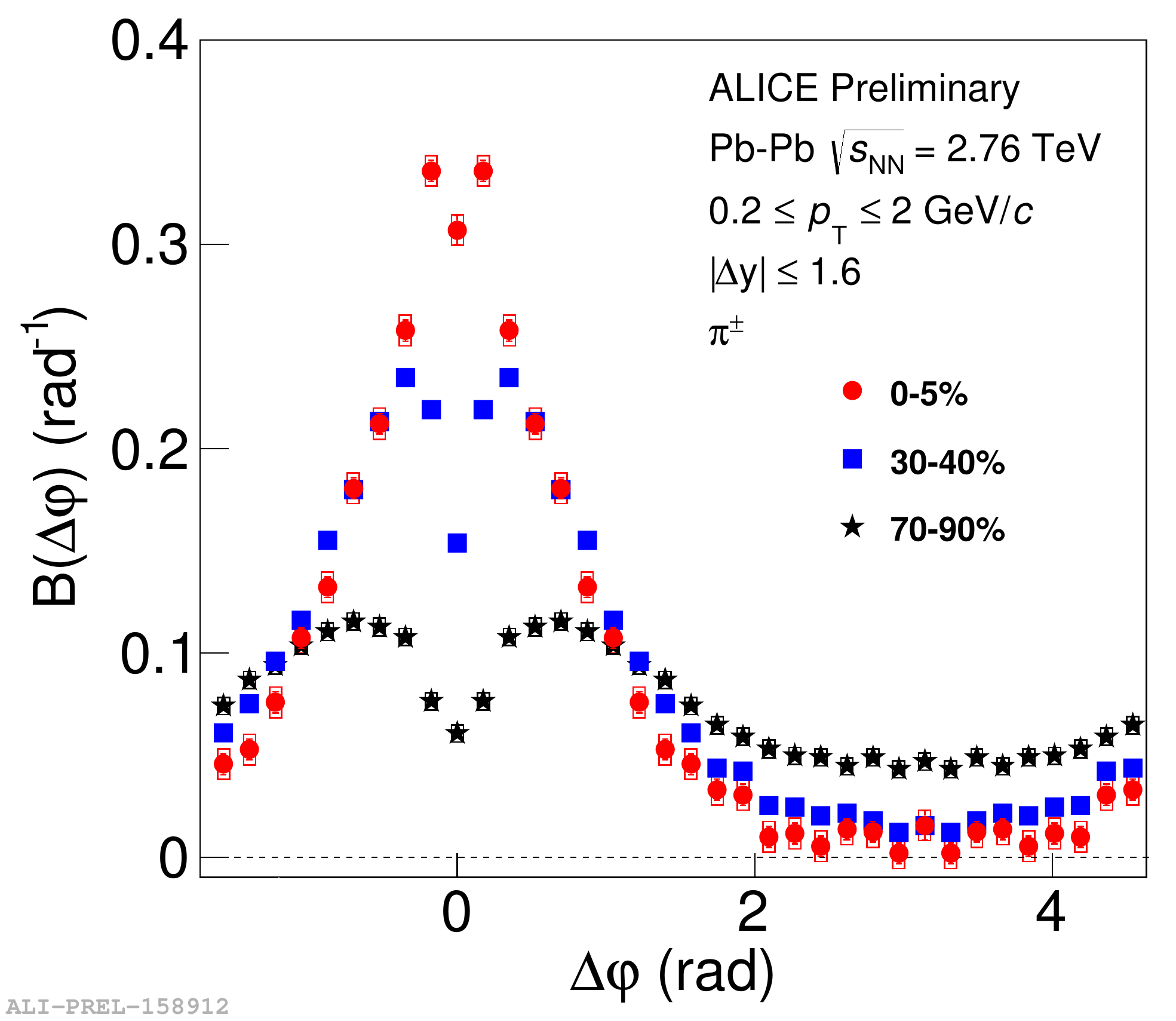}
 \includegraphics[width=0.33\linewidth]{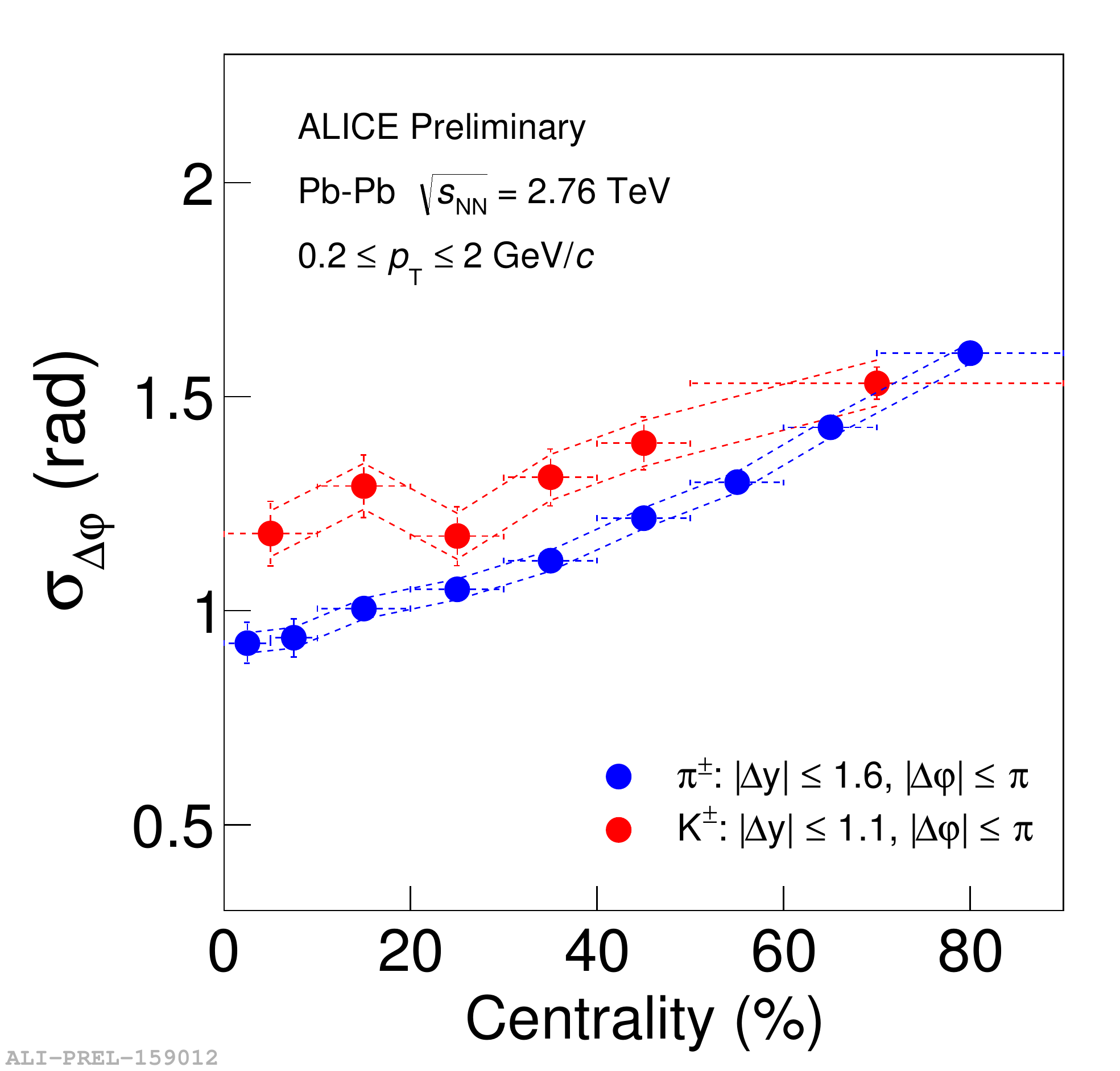}
 \caption{The BF as a function of $\Delta\varphi$ for ${\rm K}^{\pm}$ (left) and $\pi^{\pm}$ (middle) in selected centralities, and their RMS widths as a function of centrality (right) in Pb--Pb collisions at $\sqrt{s_{\mathrm{NN}}}=2.76$ TeV.}
  \label{fig:1d_PID_BF_dphi_PbPb}
\end{figure}

In this paper, only one-dimensional projections of B($\Delta y$,$\Delta\varphi$) onto $\Delta y$ and $\Delta\varphi$ axis along with their RMS widths are presented. Figures~\ref{fig:1d_PID_BF_dy_PbPb} and ~\ref{fig:1d_PID_BF_dphi_PbPb} present B($\Delta y$) and B($\Delta\varphi$) for ${\rm K}^{\pm}$ and $\pi^{\pm}$ in selected centralities, and their RMS widths as a function of centrality in Pb--Pb collisions at $\sqrt{s_{\mathrm{NN}}}=2.76$ TeV. The statistical uncertainties, usually smaller than the marker size, are represented by vertical error bars. While the systematic uncertainties, calculated as the quadratic sum of independent contributions are shown as box around each data point in the projection plots, and as dash line in the width plots. Both B($\Delta y$) and B($\Delta\varphi$) of ${\rm K}^{\pm}$ show no significant change in shape and magnitude with centrality. Note that this may in part be associated with the fact that the $\phi \to {\rm K}^{+}+{\rm K}^{-}$ decay contribution is about 30\% in the BF near-side peak as estimated from HIJING simulations [8]. The near-side is defined as $|\Delta\varphi|< \pi/2$. However, B($\Delta y$) and B($\Delta\varphi$) of $\pi^{\pm}$ show a clear evolution in shape and magnitude with centrality, with higher magnitude for more central events. There is a dip structure at $\Delta y$ $\sim$ 0 and $\Delta\varphi$ $\sim$ 0 for the BF of $\pi^{\pm}$ due to femtoscopic effects whose width grows inversely with the system size. While the dip for ${\rm K}^{\pm}$ is much less pronounced. The B($\Delta\varphi$) widths of both ${\rm K}^{\pm}$ and $\pi^{\pm}$ get narrower towards central Pb--Pb collisions with somewhat different slopes, qualitatively consistent with the presence of strong radial flow [3]. However, the B($\Delta y$) widths of ${\rm K}^{\pm}$ show no centrality dependence, while the B($\Delta y$) widths of $\pi^{\pm}$ get narrower towards central events. The different centrality dependence of B($\Delta y$) widths between ${\rm K}^{\pm}$ and $\pi^{\pm}$ in Pb--Pb collisions shows similar trends and magnitudes to STAR results in Au-Au collisions at $\sqrt{s_{\mathrm{NN}}}=200$ GeV [9], and are qualitatively consistent with the two-wave quark production model [2].  


%
\begin{figure}[h!]
\includegraphics[width=0.33\linewidth]{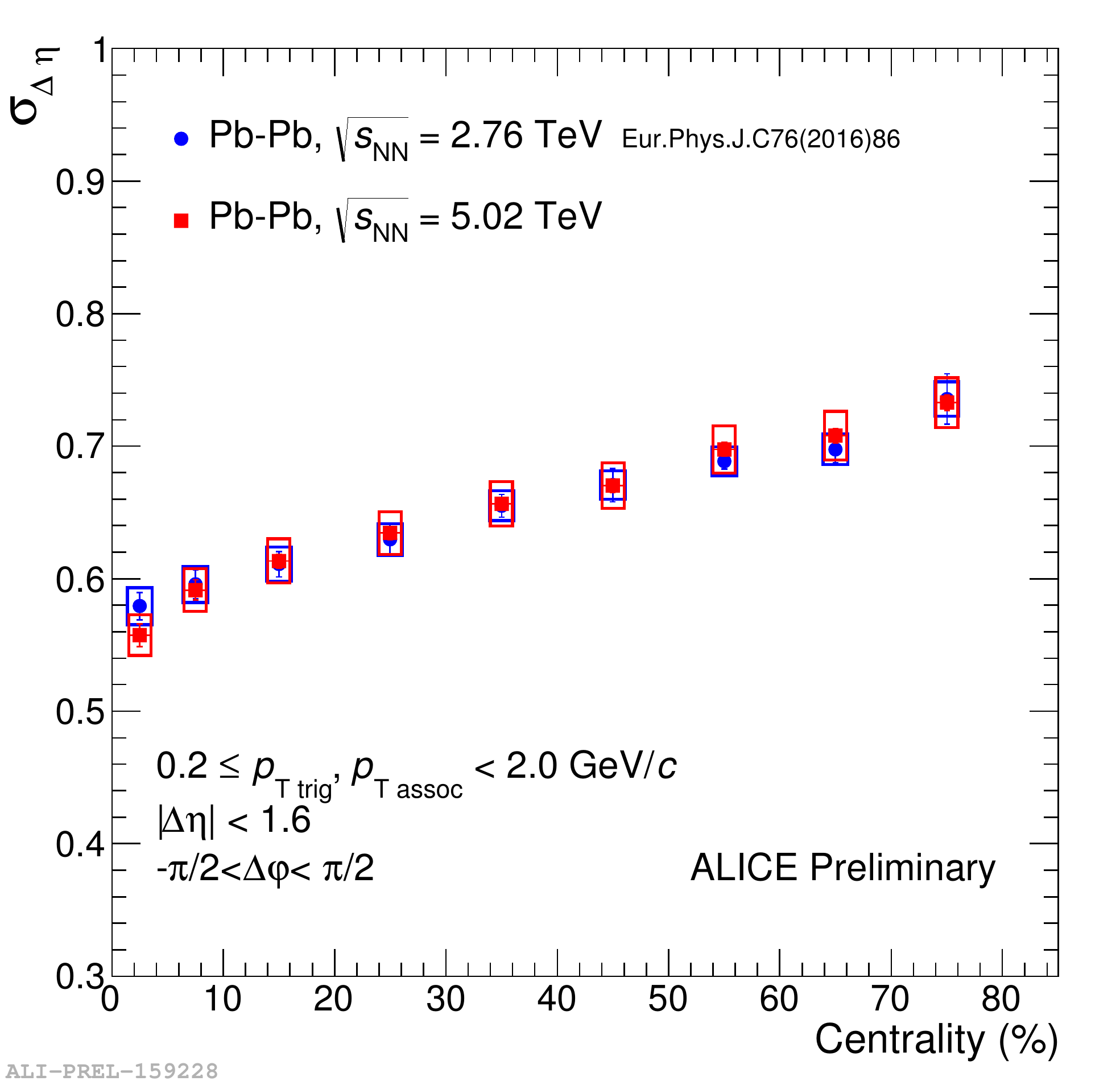}
 \includegraphics[width=0.33\linewidth]{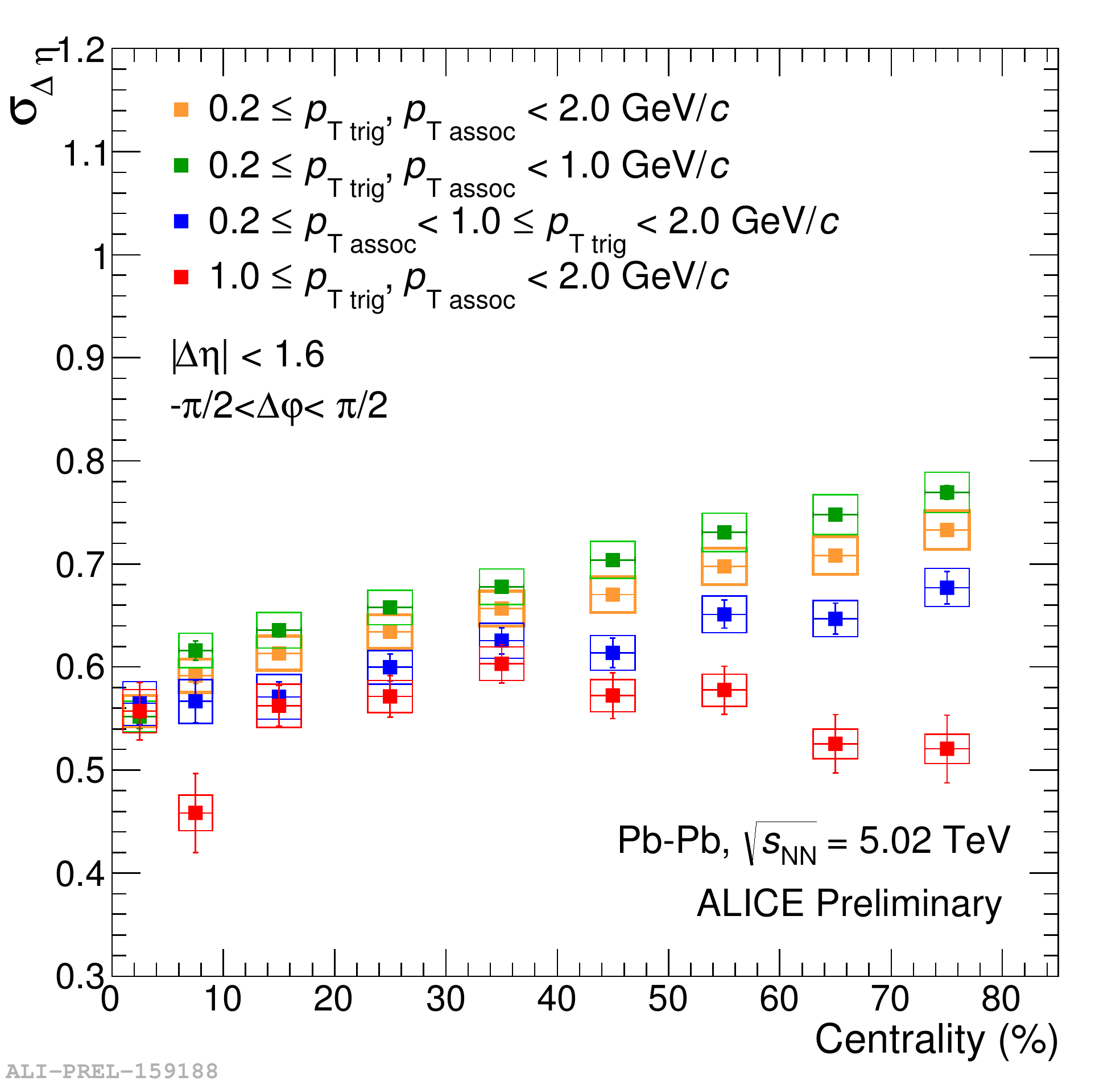}
 \includegraphics[width=0.33\linewidth]{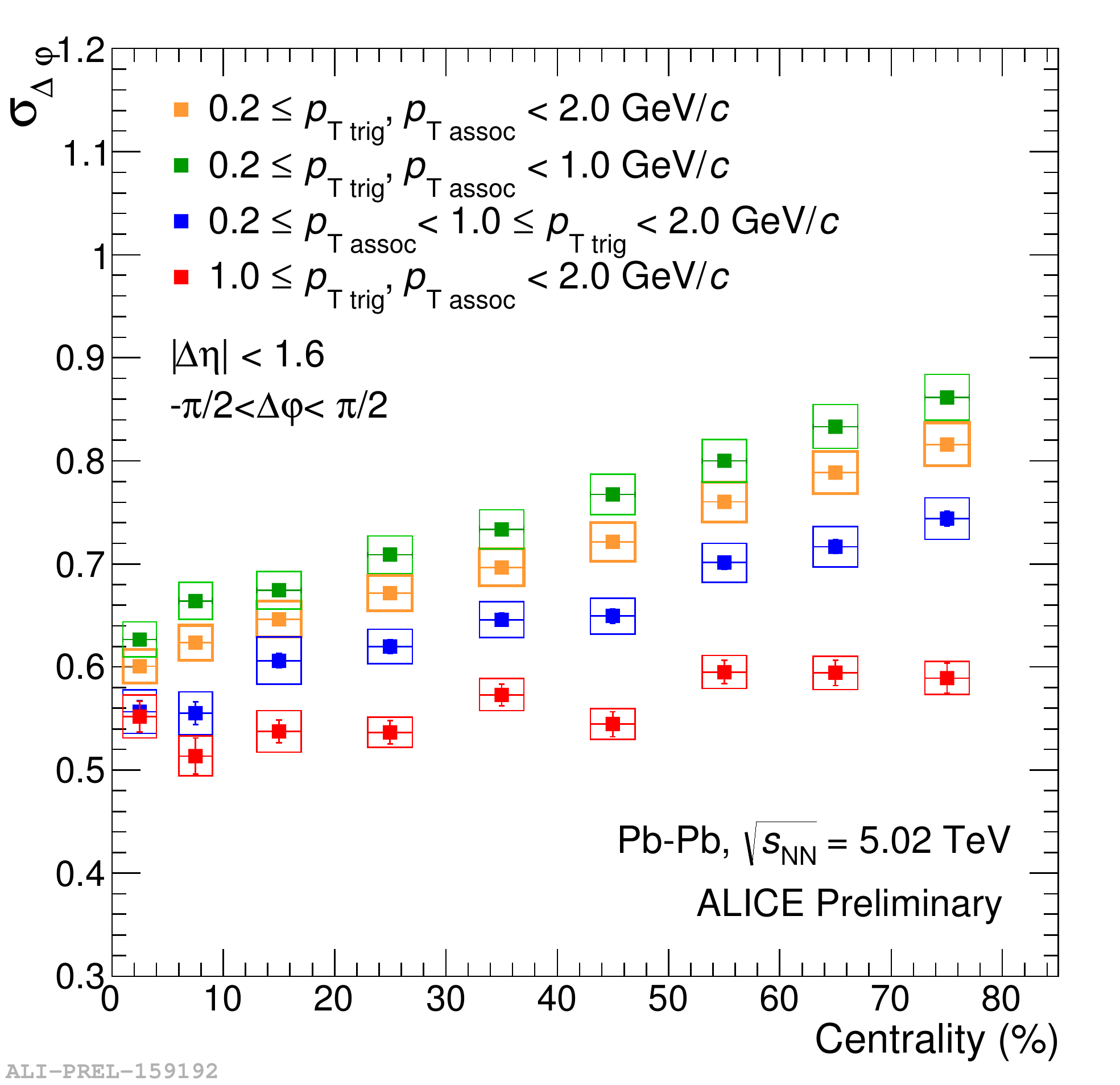}
 \caption{Centrality dependence of B($\Delta \eta$) near-side RMS widths of ${\rm h}^{\pm}$ at different collision energies (left), and B($\Delta \eta$) (middle) and B($\Delta\varphi$) (right) near-side RMS widths for different $p_{\rm T}$ ranges in Pb--Pb collisions.}
  \label{fig:Widths_Unidentified_BF_PbPb}
\end{figure}
\begin{figure}[h!]
 \includegraphics[width=0.33\linewidth]{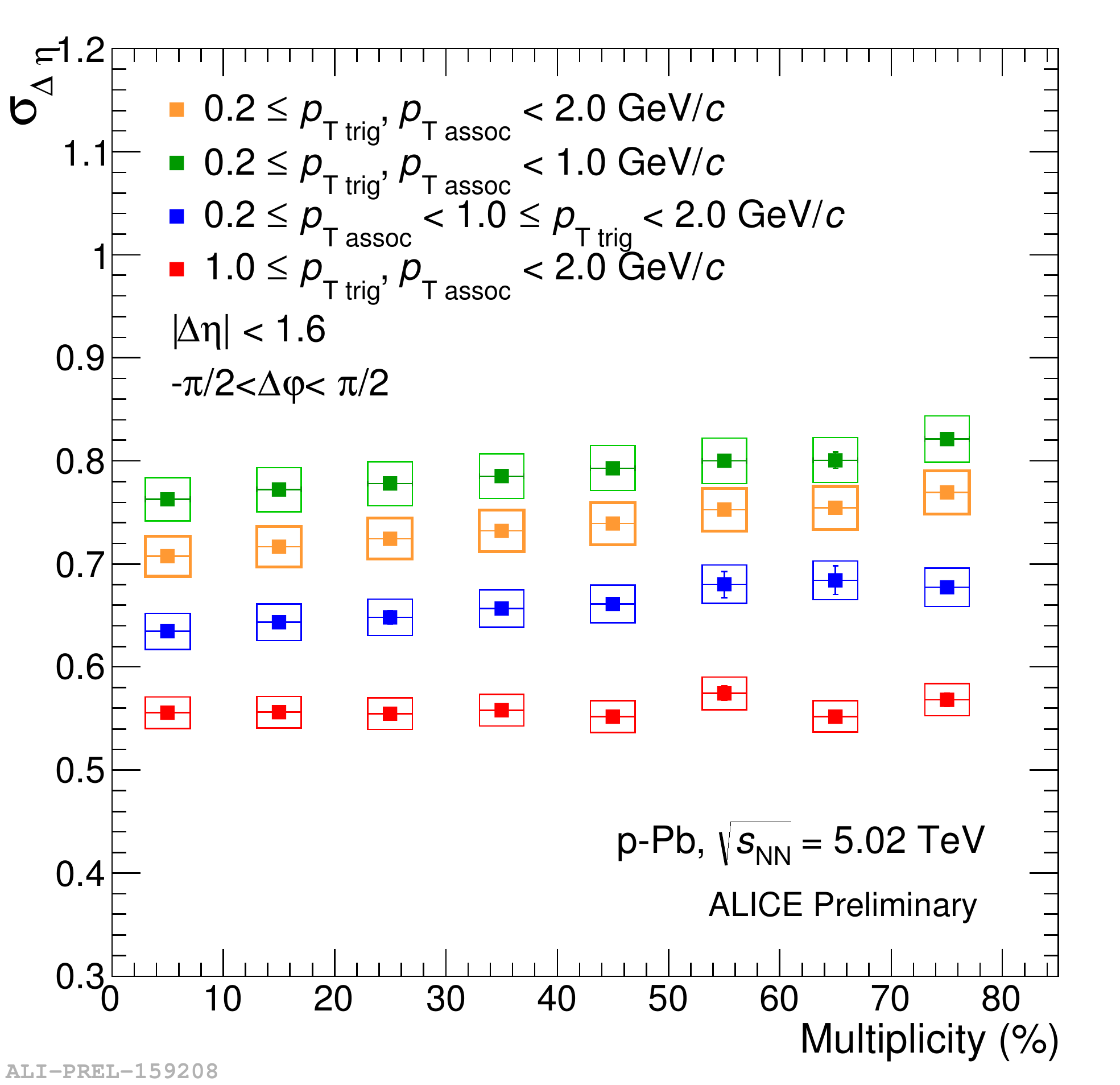}
 \includegraphics[width=0.33\linewidth]{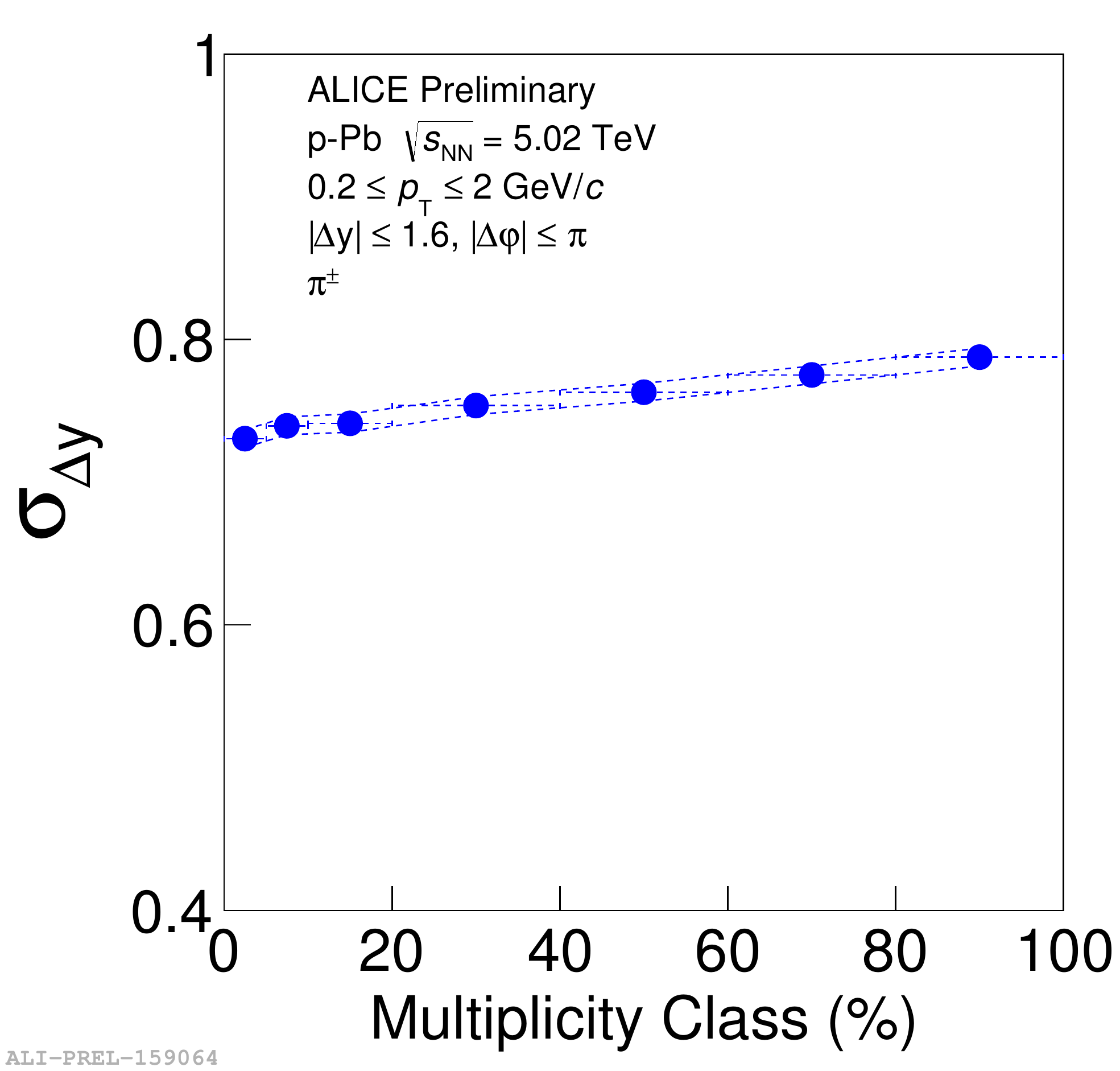}
 \includegraphics[width=0.33\linewidth]{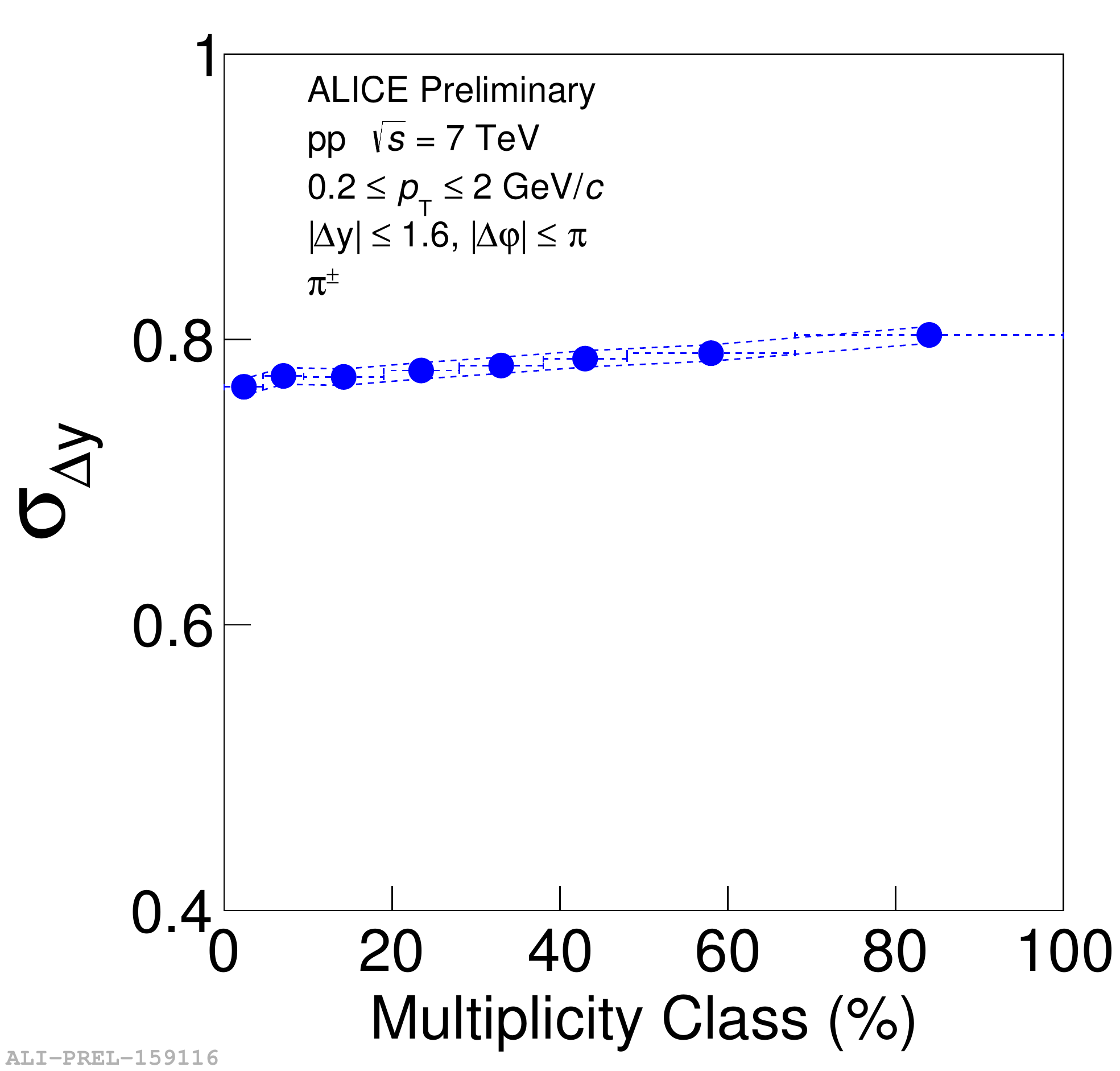}
 \caption{Multiplicity dependence of B($\Delta \eta$) near-side RMS widths of ${\rm h}^{\pm}$ for different $p_{\rm T}$ ranges in p--Pb collisions at $\sqrt{s_{\rm{NN}}}=5.02$ TeV (left), and B($\Delta y$) RMS widths of $\pi^{\pm}$ in p--Pb at $\sqrt{s_{\rm{NN}}}=5.02$ TeV (middle) and pp collisions at $\sqrt{s}=7$ TeV (right).}
  \label{fig:Widths_BF_pPb_pp}
\end{figure}

Figure~\ref{fig:Widths_Unidentified_BF_PbPb} presents that the B($\Delta \eta$) near-side RMS widths of ${\rm h}^{\pm}$ show no difference in their centrality dependence between Pb--Pb collisions at $\sqrt{s_{\rm{NN}}}=5.02$ and 2.76 TeV [10]. Additionally, the B($\Delta \eta$) and B($\Delta\varphi$) near-side RMS widths at $1.0 \le p_{\rm T trig},p_{\rm T assoc}  < 2.0$ GeV/{\it c} show no centrality dependence in Pb--Pb collisions at $\sqrt{s_{\rm{NN}}}=$ 5.02 TeV, contrary to the narrowing trends towards central events at lower $p_{\rm T}$ ranges. These results indicate that at this $p_{\rm T}$ range the initial hard parton scattering and subsequent fragmentation may already play an important role in these correlations.

Figure~\ref{fig:Widths_BF_pPb_pp} presents the multiplicity dependence of B($\Delta \eta$) near-side RMS widths of ${\rm h}^{\pm}$ at different $p_{\rm T}$ ranges in p--Pb collisions at $\sqrt{s_{\rm{NN}}}=5.02$ TeV, and B($\Delta y$) RMS widths of $\pi^{\pm}$ in p--Pb at $\sqrt{s_{\rm{NN}}}=5.02$ TeV and pp collisions at $\sqrt{s}=7$ TeV. The slopes of the RMS widths as a function of multiplicity in these small systems are smaller than in Pb--Pb collisions, which indicate different production and transport mechanisms for general charges in these small systems from Pb--Pb collisions.

%
%

%
\begin{figure}[h!]
 \includegraphics[width=0.33\linewidth]{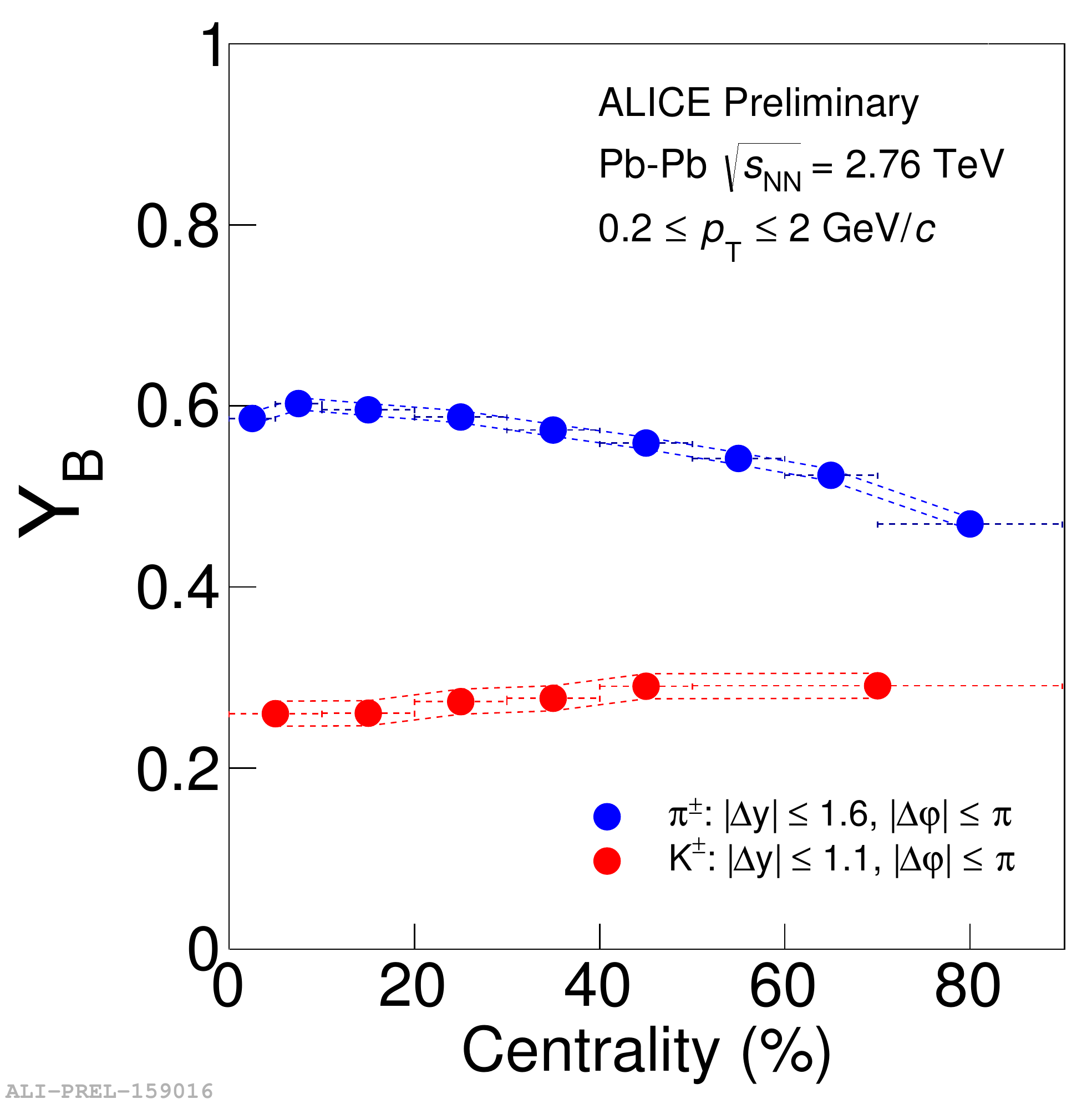}
 \includegraphics[width=0.33\linewidth]{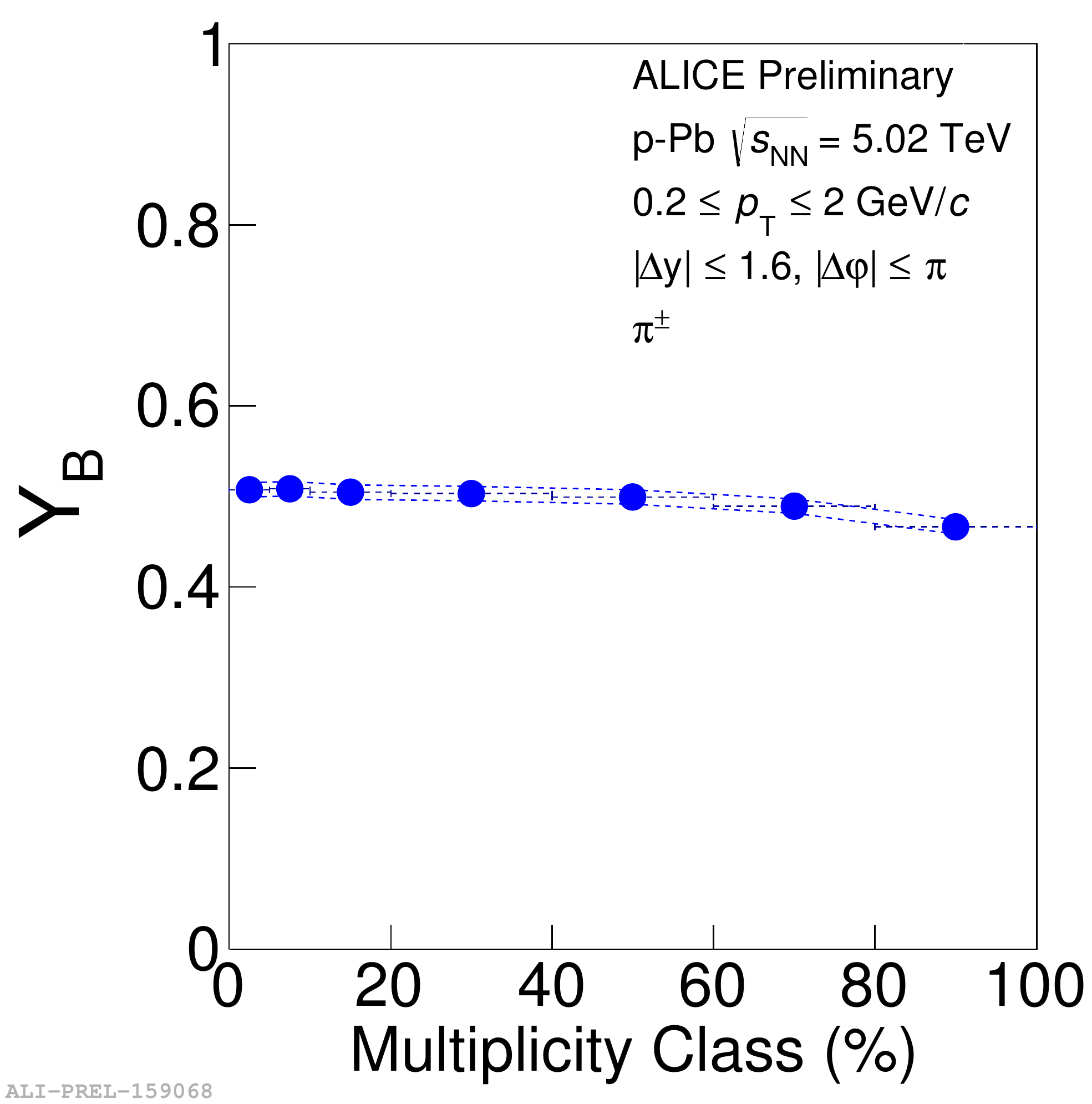}
 \includegraphics[width=0.33\linewidth]{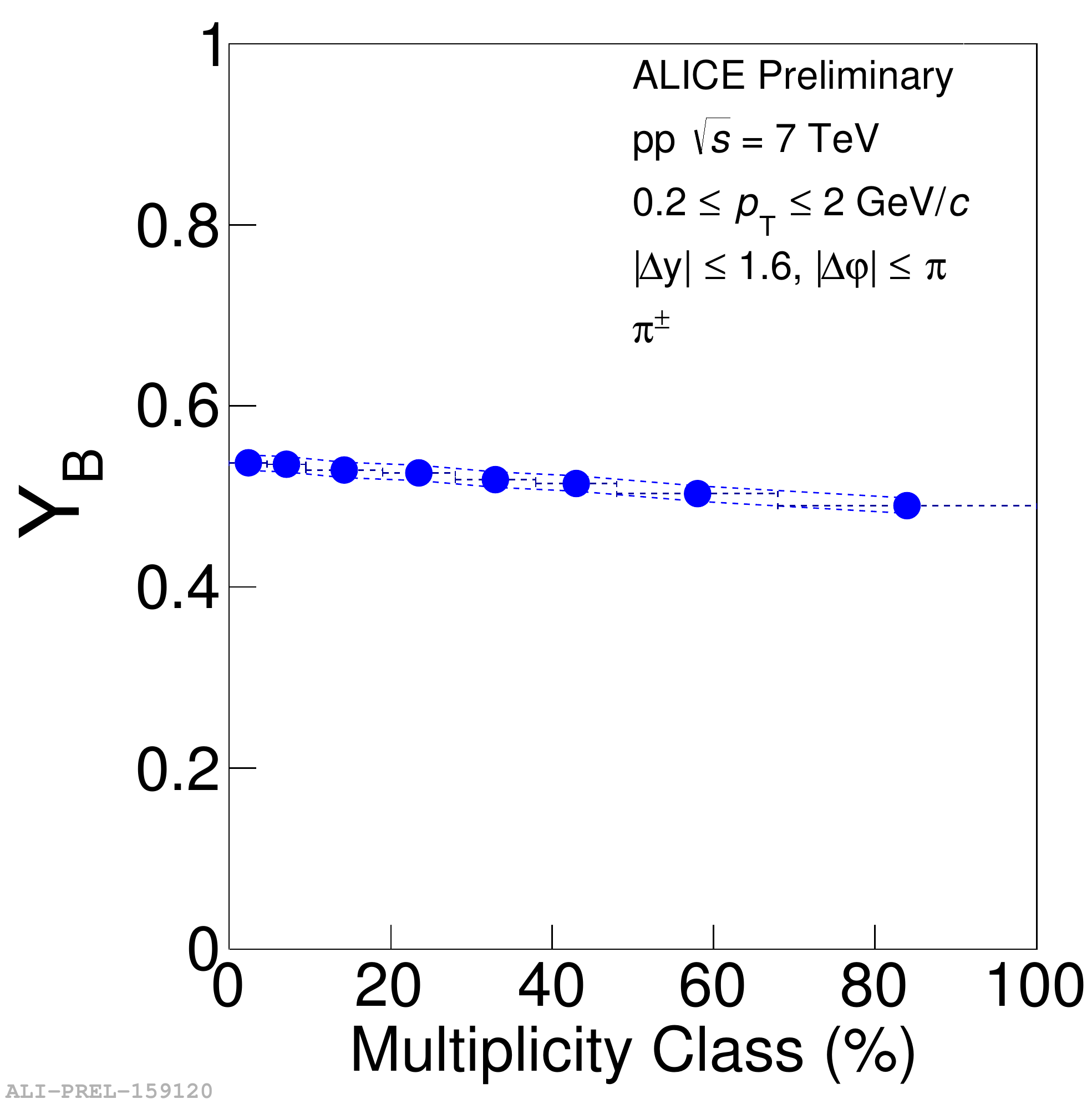}
 \caption{Centrality (multiplicity class) dependence of the BF yields for ${\rm K}^{\pm}$ and $\pi^{\pm}$ in Pb--Pb, and for $\pi^{\pm}$ in p--Pb and pp collisions.}
 \label{fig:Yields_PID_BF_PbPb_pPb_pp}
\end{figure}
%

%
%

Figure~\ref{fig:Yields_PID_BF_PbPb_pPb_pp} presents the first BF yield measurements as a function of centrality (multiplicity class) for ${\rm K}^{\pm}$ and $\pi^{\pm}$ in Pb--Pb, and for $\pi^{\pm}$ in p--Pb and pp collisions. The BF yield (or the balancing charge yield) is the integral of the BF within the acceptance. The BF yields of ${\rm K}^{\pm}$ get slightly smaller towards central Pb--Pb collisions, while the BF yields of $\pi^{\pm}$ get larger towards central (high multiplicity) events in all three collision systems with higher slope in Pb--Pb than in p--Pb and pp collisions. In addition, the centrality dependence of BF yields of ${\rm K}^{\pm}$ and $\pi^{\pm}$ in Pb--Pb collisions show opposite trends with the ALICE $K^{\pm}/\pi^{\pm}$ ratio results [11], which is probably due to the fact that correlation mechanisms and strengths are different for ${\rm K}^{\pm}$ and $\pi^{\pm}$.

\section{Summary}
\label{}

A comprehensive set of BF has been measured for ${\rm h}^{\pm}$, $\pi^{\pm}$ and ${\rm K}^{\pm}$ at various centralities (multiplicities), collision energies, and $p_{\rm T}$ ranges in Pb--Pb, p--Pb and pp collisions at the LHC using the ALICE detector. The results in Pb--Pb collisions are in qualitative agreement with expectations from the two-wave quark production model and radial flow effects. The results in p--Pb and pp collisions indicate different production and transport mechanisms for general charges in these small systems from Pb--Pb collisions.

This work has been supported in part by the Office of Nuclear Physics at the United States Department of Energy (DOE NP) under Grant No. DE-FOA-0001664.





\bibliographystyle{elsarticle-num}
\bibliography{<your-bib-database>}







\end{document}